\documentclass[aps,pre,  twocolumn,  superscriptaddress,showkeys,showpacs]{revtex4-1}
\usepackage{latexsym}
\usepackage{indentfirst}
\usepackage{graphicx}
\usepackage{color}
\usepackage{afterpage}
\usepackage{hyperref}
\def \eq #1 {\begin{equation}#1\end{equation}}
\def \eqarray #1 {\begin{eqnarray}#1\end{eqnarray}}

\def \bfE {{\bf E}}
\def \bfB {{\bf B}}

\def \bfJ {{\bf J}}

\def \wcm {\hbox{W/cm$^2$}}
\def \be  {\begin{equation}}
\def \ee  {\end{equation}}

\def \mum {\mu\mbox{m}}

\begin{document}
\title{Laser ion acceleration using a solid target coupled with a low density layer}
\author{A. Sgattoni}
\affiliation{Dipartimento di Energia, Politecnico di Milano, Via Ponzio 34/3, I-20133 Milan, Italy}\email[]{andrea.sgattoni@polimi.it}
\affiliation{Istituto Nazionale di Ottica, Consiglio Nazionale delle Ricerche (CNR/INO), research unit ``Adriano Gozzini'', Pisa, Italy}

\author{P. Londrillo}
\affiliation{INAF Bologna Osservatorio Astronomico, via Ranzani 1, I-40127 Bologna, Italy}
\affiliation{INFN sezione di Bologna, viale Berti Pichat 6/2, I-40127 Bologna, Italy}

\author{A. Macchi}
\affiliation{Istituto Nazionale di Ottica, Consiglio Nazionale delle Ricerche (CNR/INO), research unit ``Adriano Gozzini'', Pisa, Italy}
\affiliation{Department of Physics ``Enrico Fermi'', University of Pisa, Largo Bruno Pontecorvo 3, I-56127 Pisa, Italy}

\author{M. Passoni}
\affiliation{Dipartimento di Energia, Politecnico di Milano, Via Ponzio 34/3, I-20133 Milan, Italy}

\begin{abstract}
We investigate 
by particle-in-cell simulations in two and three dimensions
the laser-plasma interaction and the proton acceleration
in multilayer targets where a low
density (``{near-critical}'')
layer of a few micron thickness
is added on the illuminated side
of a thin, high density layer. 
This target design can be obtained by depositing a ``foam'' layer on a thin metallic foil.
The presence  of the near-critical plasma
strongly increases both the conversion efficiency and the energy of 
electrons and leads to enhanced acceleration of proton from a rear side
layer via the Target Normal Sheath Acceleration mechanism.
The electrons of the foam are strongly accelerated in the forward direction and 
propagate on the rear side of the target building up a high electric field
with a relatively flat longitudinal profile.
In these conditions the maximum proton energy is up to three times higher 
than in the case of the bare solid target. 
\end{abstract}
\maketitle

\section{Introduction}
The availability
of laser systems with short and high intensity pulses allowed the 
development of different techniques for the acceleration of both 
electrons \cite{gibbon-book,esareyRMP09}
and ions \cite{borghesiFST06,macchiRMP11}
exploiting the
laser-plasma interaction and the high electric field generated in the plasma.
The plasma is usually obtained from the ionisation of the chosen target
by the laser pulse itself and, depending on the material, different values of
the  plasma density can be achieved leading to different interaction regimes.
In the interaction of an electromagnetic (EM) wave 
of frequency $\omega$ (i.e. a laser pulse) 
with a plasma, the values of the electron density $n_e$ and of the parameter 
$n_c=m_e\omega^2/4\pi e^2$ (so-called \emph{critical} or cut-off density)
discriminate between two different regimes. For \emph{underdense}
plasmas where $n_e<n_c$, according to linear theory the laser pulse propagates
into the plasma and thus the interaction occurs through the entire plasma 
volume. Collisionless absorption of the laser energy may occur, e.g., via the 
excitation of plasma waves due to wakefield generation or parametric 
processes (Raman scattering) \cite{esareyRMP09}. 
In turn, the plasma waves may break and lead
to the generation of fast electrons, whose energy and number increase with the
plasma density \cite{modenaN95}. 
For \emph{overdense} plasmas where $n_e>n_c$, the 
laser only penetrates in the ``skin'' layer of thickness 
$\sim c/\omega_p=(\lambda/2\pi)\sqrt{n_c/n_e}$ 
and thus a surface interaction, rather than a volume 
interaction occurs. A considerable amount of the laser energy is usually 
reflected, but a sizable
or even major part may be absorbed via resonant or 
non-resonant excitation of surface plasma oscillations which, in a steep 
profile, may also break and generate fast electrons 
\cite{gibbon-book}.
In this regime usually the absorption fraction increases with decreasing 
density as the laser pulse penetrates more deeply into the plasma. 

In the context of experiments using high-intensity lasers with optical or
near-infrared wavelengths ($\lambda=0.8-1\mum$), 
most of the times the underdense and overdense
regimes correspond in practice to the use of gas targets with densities 
$n_e\ll n_c$ and of solid targets with $n_e\gg n_c$, respectively.
A regime which is at the boundary between underdense
and overdense plasma may lead to efficient absorption and fast electron generation,
according to the observed scaling with 
density.
Despite the interest in such ``intermediate'' conditions with
$n_e \simeq n_c$ \cite{WillingalePRL06,BulanovPRL07,bulanovjrPoP10},
these conditions have not been accurately investigated with experiments
because such targets are not straightforward to produce.
Actually, in real experiments laser prepulses often lead to early plasma
formation from solid targets so that the short-pulse, high intensity 
interaction actually occurs with an inhomogeneous plasma with both 
underdense and overdense regions. However, such conditions are usually out
of a complete experimental control and both interpreting experimental data
and optimising energy absorption are not straightforward. 
This issue is of particular relevance for proton acceleration via the 
Target Normal Sheath Acceleration (TNSA) mechanism, where protons on the
rear side of the target are accelerated by the space-charge electric field 
generated by fast electrons having crossed the target and escaping in vacuum.
Thus, increasing the conversion efficiency and the energy of electrons 
generated at the front side yields a more efficient TNSA of more energetic
protons as well. 

Some recent experiments aimed at improving ion acceleration by laser 
pulses
have considered solid targets of thickness short enough that 
the long laser prepulse produces a near-critical plasma \cite{yogoPRE08},
or even ultrashort thickness such that the
expansion during the interaction leads to self-induced transparency, i.e.
to an overdense to underdense transition \cite{henigPRL09a,jungPRL11}. 
Rather thick ($> 10^2~\mu\mbox{m}$) \emph{foam} targets 
with low density
have been used in experiments either at relatively moderate
($\sim 10^{19}\mbox{W cm}^{-2}$) \cite{liPRE05}
or at very high intensity ($\sim 10^{21}\mbox{W cm}^{-2}$) 
\cite{willingalePRL09,willingalePoP11}, 
and relatively ``long'' 
($> 5 \times 10^2\mbox{fs})$ pulse {durations}. 
Thinner foams 
and ultrashort pulse durations (tens of fs)
have been studied with particle-in-cell (PIC) simulations both 
for intensities in the moderate 
($\sim 10^{19}\mbox{W cm}^{-2}$) \cite{nakamuraPoP10} and
and ultra-high ($\sim 10^{23}\mbox{W cm}^{-2}$) \cite{bulanovjrPoP10} regimes.
Most of the above mentioned studied envision ion acceleration
mechanism which may be considerably different from TNSA. However, TNSA is the 
mostly investigated mechanism so far, and its peculiar features such as 
broad spectrum and ultra-low emittance proton emission make it most suitable
for specific important applications such as time-resolved proton radiography 
\cite{gibbon-book,borghesiPoP02}. 
It is thus of interest to couple the potential of low-density
targets with the TNSA scheme.

The typical proton acceleration scheme in the TNSA regime consists of 
a thin solid foil ($l\simeq 1-10\mu\mbox{m}$)
irradiated by a high intensity laser ($I=10^{19}-10^{21}\mbox{W/cm}^2$).
The maximum proton energy is mainly limited
by the laser energy absorbed by the electrons.
Material science offers opportunities for manufacturing targets with low
and controlled density, such as foam targets. The possibility to couple
a thin solid foil with a near critical density layer attached on the 
irradiated side can enhance the laser energy absorbed by the
target with a consequent increase of the electron energy and the accelerating
field arising on the rear side of the target. An ``advanced'' TNSA regime
may be obtained with a higher proton energy considering the same laser
characteristics.

Here we report a simulation study of laser
absorption and ion acceleration for a three-layer target configuration, 
where a low density, few-microns thick (``foam'') layer 
and an ultrathin (``impurity'') proton layer are deposited on the 
front and rear sides, respectively, of a (``solid'') foil target.
A similar three-layer target configuration has been studied by \textcite{nakamuraPoP10} using 2D simulations at normal
incidence. An enhancement of proton acceleration due to the presence of the foam has been shown together with
the effects of the field ionisation.
In the present work we extend the investigation in different directions.
We present fully three dimensional (3D) simulations and 2D simulations with oblique incidence.
We use the PIC code ALaDyn \cite{BenedettiIEEE08,LondrilloNIMA2010}, 
in both 2D and 3D, to investigate the dynamics of the laser interaction with the
slightly overcritical density plasma and the role of the foam electrons in the
rise of the longitudinal electric field which accelerates the protons.
Both 3D and 2D simulations show the proton
energy ``gain'' obtained considering a target with a foam layer
instead of a ``bare'' target with no foam. 
We show the importance of a 3D study to correctly determine the maximum value of the
proton energy and the dynamics of the acceleration. A wide parametrical study in 2D as a
function of foam thickness and density and of the incidence angle of the laser is also reported.
We quantitatively analyze the absorption of the laser energy by the foam-solid system, considering
different parameters of the target and relating it to the energy of the accelerated protons.
We address in detail an important aspect of the proton acceleration mechanism, namely the
formation of a strongly “non-equilibrium” electric field at the rear side of the target,
due to the prompt escape in vacuum of bunches of highly relativistic electrons created
in the laser foam interaction.

\section{Simulation set-up}
In the PIC simulations we consider the case of a 
$p$-polarised laser pulse incident
(at normal incidence) on an already ionised, ideal plasma.
The target is composed by three layers: a thin metallic
foil ($\sim0.5\mum$) (main layer), a thicker low density foam ($1-12\mum$)
attached  on the side directly irradiated by the laser beam, and a third layer
representing the thin ($\sim 50\mbox{nm}$) contaminants layer on the rear surface.
The ionization state of the target layers is fixed and the
charge over mass ratio are
$Z/A=1/2$, $Z/A=1/3$ and  $Z/A=1$ for the foam layer (e.g. C$^{6+}$),
metal foil (e.g. $\mbox{Al}^{9+}$) and contaminants ($\mbox{H}^+$) respectively. 
The laser propagates along the $x$
direction and in every case considered 
independently on the foam thickness, it reaches the maximum focusing
on the front surface of the  metal foil.
In all the simulations reported in the paper
the laser pulse wavelength was fixed to $\lambda=0.8\mum$ and the 
pulse shape was kept as $\sin^2$ and Gaussian in the longitudinal and the 
transverse direction respectively
with $\tau=25$fs and $w_0=3\mu$m being the time duration (FWHM) 
and pulse waist.
The peak value of the dimensionless amplitude of the laser pulse 
$a_0=8.5\cdot10^{-10}\lambda[\mum]I^{1/2}[\wcm]$ was varied in the range 
$a_0=3-20$, which corresponds to a  
peak power range $P=2.8-128\mbox{TW}$ and 
total energy $U=0.075-3.2\mbox{J}$, well within the capabilities of many present-day facilities. 
The spatial resolution is  $\Delta x=c/2\omega_p\simeq \lambda/111$ $\Delta y=\Delta z=c/\omega_p$, where $\omega_p$ is
evaluated from the maximum density of the plasma (i.e. the solid foil density).

Two 3D simulations have been performed considering the same solid target with and without
a foam layer  $n_f=2n_c$, $l_f=2\mu\mbox{m}$.
In these cases the ``contaminant layer'' and metallic foil 
density and thickness are $n_r=9n_c$, $l_r=0.05\mu$m and
$n_m=40n_c$, $l_m=0.5\mu$m respectively. 
The  electron population of the
main foil and of the foam layer have  been sampled with respectively
36 and 8 macro-particles/cell, whereas the ion populations with 12 and 8 macro-particle/cell;
 protons and electrons of the contaminants layer are sampled with 36 macro-particles/cell.
The total number of grid points of the 3D simulations
has been about $3\cdot10^9$ and the number of macro-particles
is about $6\cdot10^9$; these simulations required about 30000 CPU-hours
on the SP6 machine of CINECA (Bologna, Italy).

A more extensive investigation spanning on a wider range of foam parameters
has been performed with several 2D simulations.
The electron density of the foam $n_f$ and its thickness $l_f$ 
have been varied:  $n_f=1-8n_c$, $l_f=1-12\mu$m, whereas
the main layer density has been fixed to $n_m=80n_c$.
The number of macro-electrons per cell was 81/cell in the solid density layer, 9 for the contaminants
layer and 16 or 25 for the foam layer, whereas the number of macro-ions of the three species was 9/cell.

The time $t=0$ corresponds to the
instant when the laser starts to interact with the plasma;
most of the simulations have been stopped after $50\mu\mbox{m}/c=166\mbox{ fs}$.
A comparison of the results obtained in 2D and 3D shows
how in 3D the acceleration process at $t=166\mbox{ fs}$ 
is almost finished, whereas in the
2D simulations the acceleration process is still significant even after nearly 400fs.
In a 2D simulations the ``point'' charges effectively behave as infinite wires with finite
linear charge density. On longer time-scales the problem is nearly pure electrostatic and the logarithmic behaviour of the
potential ($\phi\sim-\ln(r)$) leads to a much slower ``saturation'' of the acceleration process and higher proton energies.

\section{Results and discussion}

\subsection{3D simulations}
The density of the solid foil chosen for the 3D simulation is $n_m=40n_c$;
though less than the value for a real solid metal (several hundreds $n_c$),
it is considerably higher than the critical density and, for a peak
intensity $a_0=10$, it ensures that the plasma is overdense even accounting
for relativistic transparency effects \cite{cattaniPRE00}.
A set of 2D simulations strengthened the choice of the foil density:
we considered cases varying the laser intensity  ($a_0=10-20$)
and the plasma density of the main foil
($n_m=40\;\mbox{and}\;80n_c$).
For the lower laser intensity ($a_0=10$), when no foam layer is present the
results obtained with the two different main target densities ($n_m=40\;\mbox{or}\;80n_c$) are similar,
(proton maximum energy is 11 instead of 9 MeV) but a lower density, as expected, leads to better absorption,
whereas they are barely distinguishable in presence of a foam.
For the higher value of the pulse intensity  ($a_0=20$), the difference
between the case of $n_m=40n_c$ or $n_m=80n_c$ is more significant with the lower density
leading to considerably higher proton energy and enhanced laser energy absorption in both cases, with and without foam.
This can be explained by the fact that for the high intensity case
the foam layer considered is too thin to absorbs a sizable fraction of the total laser
energy, thus the interaction mostly occurs with the solid foil and is 
more sensitive to the density of the latter.
Additional simulations
have been performed doubling the thickness of the solid foil ($n_m=80n_c$, $l_m=1\mum$ instead of $l_m=0.5\mum$).
The maximum proton energy in presence of the foam layer ($l_f=4\mum$ $n_f=1n_c$) is reduced by a few percent ($\sim7\%$, 24 instead of 26 MeV),
whereas for a bare target, as also reported in previous work \cite{esirkepovPRL06},
it decreases more substantially ($\sim25\%$, 6.5 instead of 9 MeV).
It can be concluded that if the foam thickness 
is sufficient to absorb a major part of the laser energy, the characteristics of the solid foil are less crucial.

\begin{figure}[tb]
\begin{center}
\includegraphics[width=.4\textwidth]{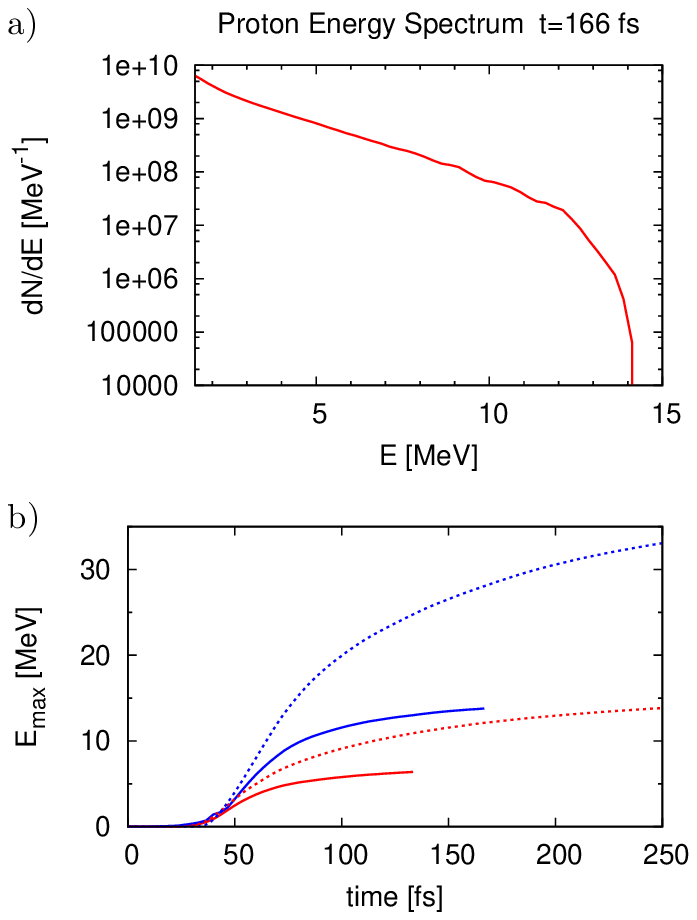}
\end{center}
 \caption{a): proton energy spectrum at $t=166$ fs considering a laser pulse with $a_0=10$ and a target with a foam
layer $n_f=2n_c$, $l_f=2\mu$.
b): proton maximum energy evolution with respect to time: comparison 3D (solid) and 2D (dashed) cases
without foam (red) and with (red)  $n_f=2n_c$, $l_f=2\mu$, $n_m=40n_c$, $a_0=10$.
}
 \label{fig: 3D Pasquale}
 \end{figure}

The simulations  led to
proton energy spectra
with an exponential profile and a cut-off typical of the TNSA regime (Figure \ref{fig: 3D Pasquale}).
The energy spectra are obtained considering all the protons of the contaminants layer
achieving a spatial integration in the transverse direction.
The presence of a collimator after the target would then select the high
energy tail of the spectrum similarly to the TNSA case.
At a laser intensity corresponding to $a_0=10$, in presence of
a foam layer, although rather thin ($2\mu\mbox{m}$), the maximum proton energy $E_{max}$ is much higher,
 $E_{max,f}\simeq14\mbox{MeV}$  than without foam $E_{max,b}\simeq6\mbox{MeV}$ (see fig \ref{fig: 3D Pasquale}).
The same simulations have been performed
also in 2D:  the cut-off energy of the protons is overestimated (by a factor of about 2 at time $166$fs)
but the ratio $E^{2D}_{max,f}/E^{2D}_{max,b}\simeq E^{3D}_{max,f}/E^{3D}_{max,b}\simeq 2.3$
is preserved. 
It is therefore evident that the 3D analysis is essential to evaluate the
maximum ion energy quantitatively. Nevertheless, the observed dynamics of the 
laser-foam interaction is qualitatively similar in 2D simulations, thus we use 
the latter for a parametric study aimed at both showing the existence of an 
optimal foam  thickness as a function of the laser
amplitude and angle of incidence, and at 
evidencing features of the electron acceleration and sheath field formation 
processes (Sec.\ref{secIIIB}).

\begin{figure}[t]
\begin{center}
\includegraphics[width=.4\textwidth]{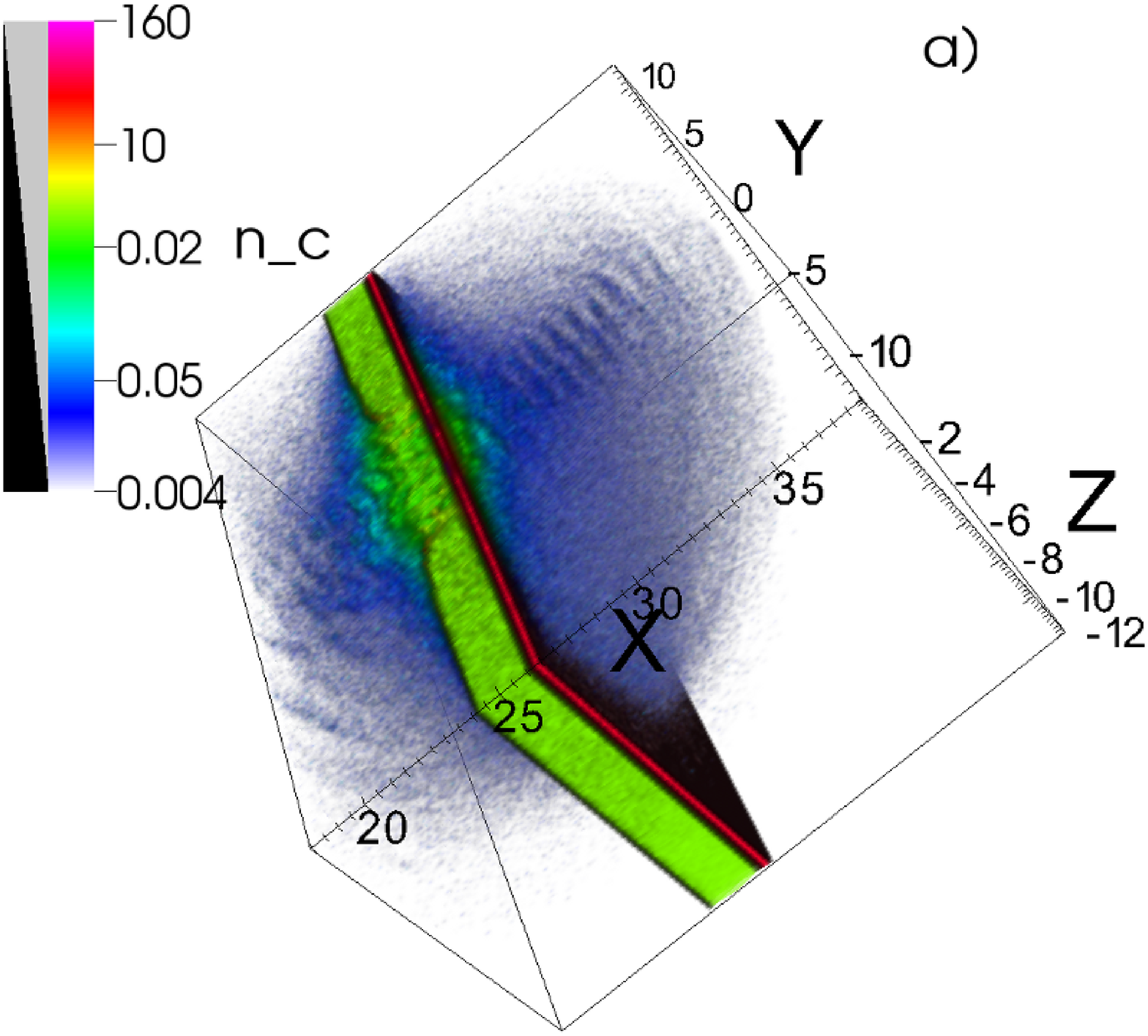}
\includegraphics[width=.4\textwidth]{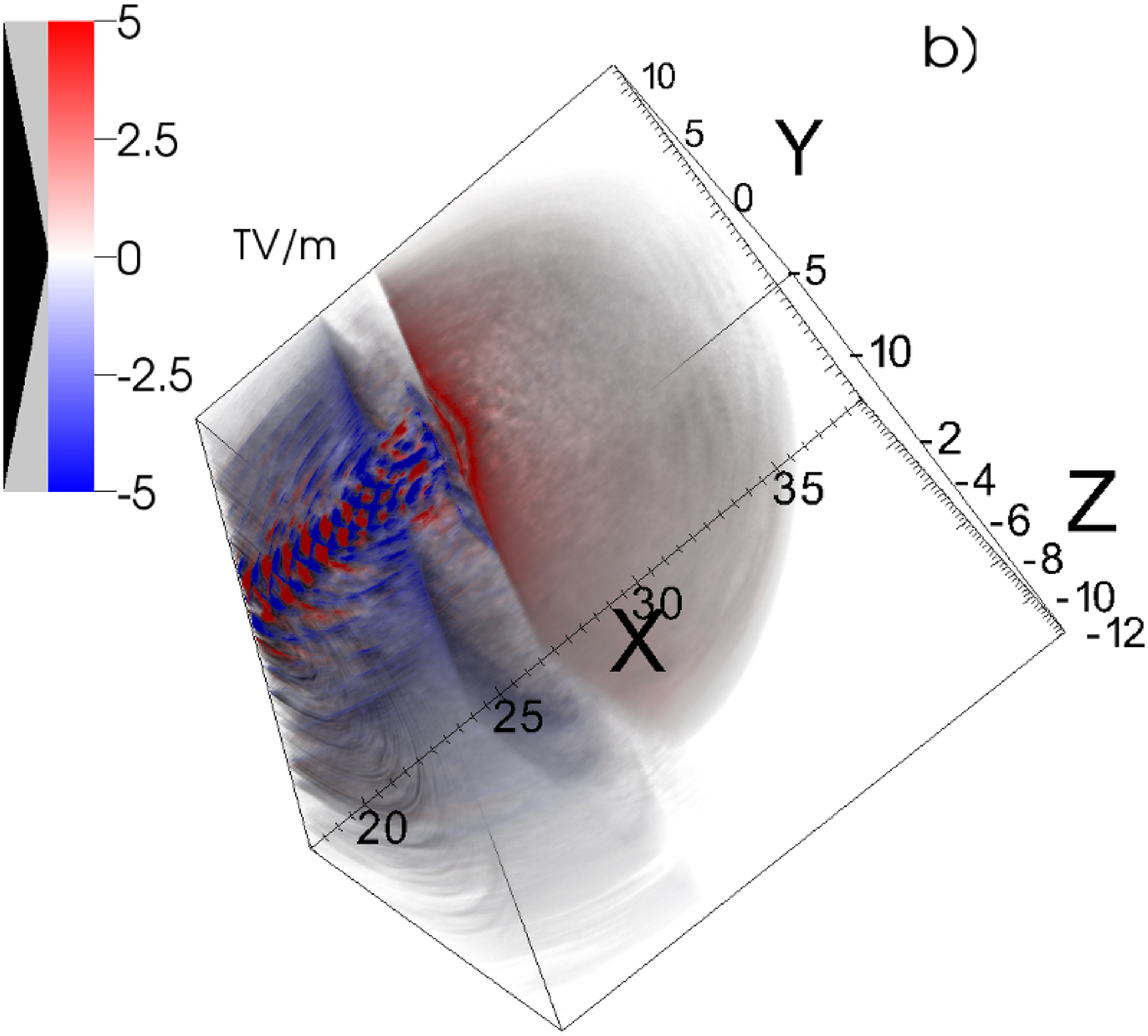}
\end{center}
 \caption{3D simulations ($a_0=10$, $w_0=3\mum$, $l_f=2\mum$, $n_f=2n_c$):
electron density in logarithmic colour scale (a)
and longitudinal electric field (b) at $t=66\mbox{ fs}$ in presence of the foam.
Half of the simulation space has been removed ($z>0$) for easier viewing.}
 \label{fig: 3D3D}
 \end{figure}

In Figure \ref{fig: 3D3D} the volume rendering of the electron density is presented together with
the distribution of the longitudinal electric field. The electrons from the
foam are  accelerated in the forward direction and reach the rear side of
the target. The electron cloud expands with a roughly spherical
symmetry for several microns and displays regular structures in the
longitudinal direction which are more evident in the
2D plots and will be further discussed.
The resulting charge separation leads to a strong longitudinal electric field
which exceeds $5\mbox{ TV/m}$ and extends in the longitudinal direction according to
the electron distribution for about 10 microns.
The slice displayed in the Figure \ref{fig: 3D3D} highlights the longitudinal electric field of
the laser pulse and shows how the laser reaches the solid foil and is then reflected by the high density plasma.
Thus, even if $n_e>n_c$ the foam plasma is effectively underdense for the 
laser pulse, due to relativistic penetration and ponderomotive channeling effects.
Since an extended analysis of the laser plasma interaction and ion acceleration
 for several values of the parameters
can be hardly performed in 3D we continued the investigation with
2D simulations.

\subsection{2D simulations}
\label{secIIIB}
The detailed analysis of the results for several target parameters has been
carried on the basis of 2D simulations and the discussion will now focus on 2D cases only.
In the light of the 3D results and the comparison with the analogous 2D cases,
we will focus the attention on the main features of the laser-plasma interaction as well of
the foam target parameter rather than on a quantitative estimation of the proton energies.

\begin{figure}[tb]
\begin{center}
\includegraphics[width=.4\textwidth]{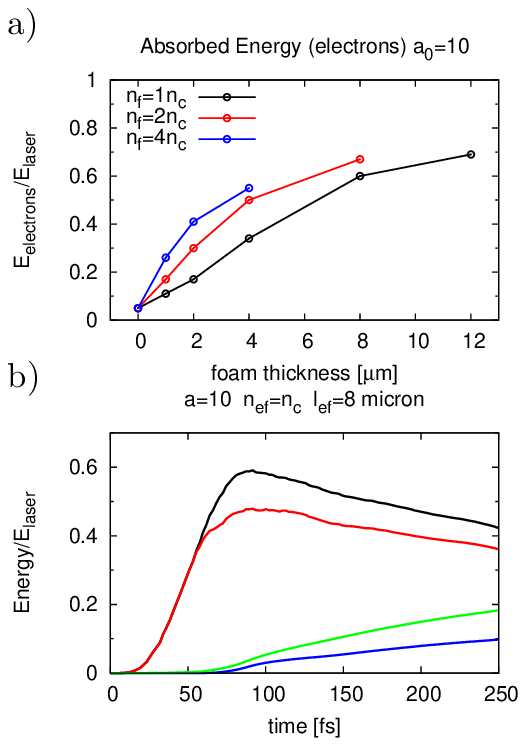}
\end{center}
\caption{a): maximum value of the electron energy absorbed by the target for different cases.
b): for the case with $l_f=8\mu$m:
the time evolution of the energy normalised to the initial laser energy of:
 all electrons (black), foam electrons only (red), all ions (green), contaminant protons only (blue). }
\label{fig: energy evol foam n1}
\end{figure}

When a low density foam layer is present, the energy absorption
mechanism is different from the case of a highly overdense plasma:
in the range of parameters here considered
the laser propagates through the
foam and is not effectively reflected until it reaches the solid layer.
A minor part of laser energy is then
absorbed by the high density plasma similarly to the case without foam.
During the interaction with the foam, the laser pulse accelerates the electrons
to relativistic velocities and its energy is considerably depleted.
We analysed the energy balance of the EM energy and 
kinetic energy of the particles (Figure \ref{fig: energy evol foam n1}).
The parameter scan in 2D, Figure \ref{fig: energy evol foam n1}-a, shows that
with the increase of the foam thickness ($l_f$), for a constant foam density,
 the ratio of the energy absorbed by the electrons over the
initial laser pulse energy increases.
For a constant foam density $n_f=n_c$, the estimated reflectivity
is reduced from 92\% for the case of ``bare'' solid target,
to 18\% using a $8\mu$m foam.
The right panel of Figure \ref{fig: energy evol foam n1} shows the time evolution
of the particles kinetic energy for a case with a foam layer
with thickness $l_f=8\mum$ and density $n_f=n_c$.
When a sufficiently thick foam is present ($l_f\geq 4\mu\mbox{m}$), the electrons gain a notably large
fraction of the initial laser energy (up to $>50\%$ of the total energy
whereas without foam the corresponding value
is about $5\%$).
The ions slowly gain energy at the expense of
both the electron kinetic energy and, to a lower extent,
the electrostatic energy. 
It is remarkable how the protons from the
contaminants, although forming a thin (50 nm) and low density ($n_r=9n_c$) layer,
get about 50\% of the total kinetic energy absorbed by
all the ion population of the simulation, 
accounting for up to 10\% of the initial laser energy (to be compared with
$\simeq 1\%$ of the corresponding case without foam).

\begin{figure}[tb]
\begin{center}
\includegraphics[width=.4\textwidth]{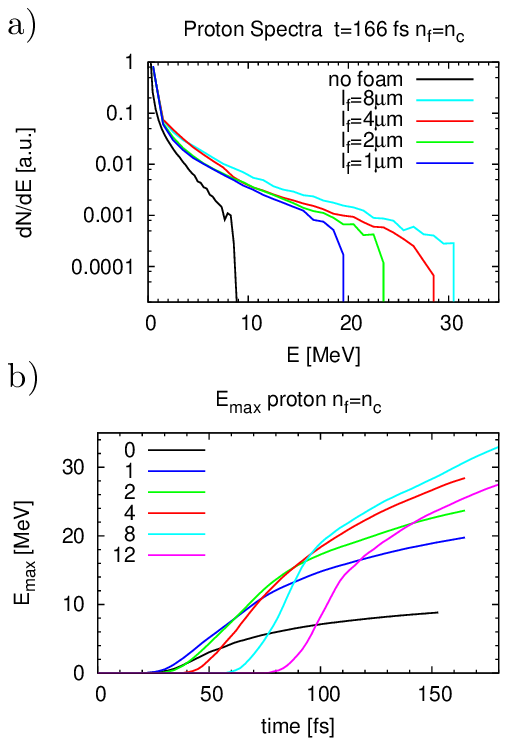}
\end{center}
 \caption{a): proton energy spectra at  $t=166\mbox{ fs}/$ considering a laser with $a_0=10$
and foam layer with density $n_f=n_c$.
b): maximum proton energy evolution with respect with time for different
foam thicknesses ($a_0=10$, $n_f=n_c$).}
\label{fig: spectra evolution}
 \end{figure}

\begin{figure}[b]
\begin{center}
\includegraphics[width=.4\textwidth]{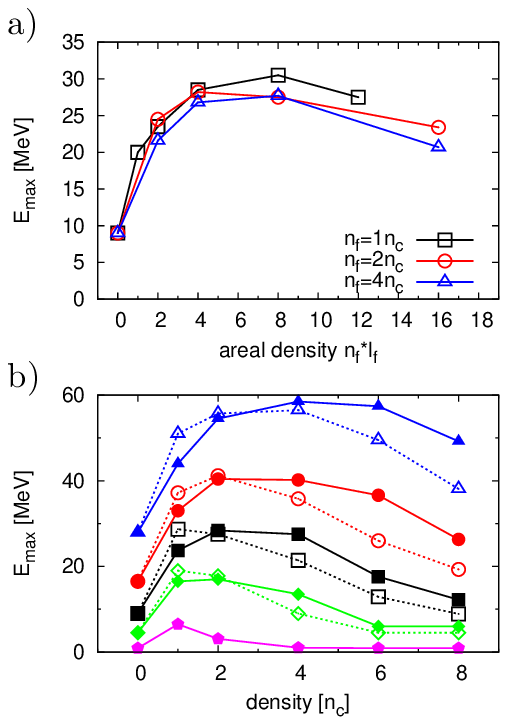}
\end{center}
 \caption{a): Proton maximum energy for $a_0=10$, different foam thicknesses and foam densities
(black squares: $n_f=n_c$, red circles: $n_f=2n_c$, blue triangles: $n_f=4n_c$)
as function of the areal density ($l_fn_f/n_c$).
b): Proton maximum energy versus foam density for different thickness (2 $\mu\mbox{m}$ solid line, 4 $\mu\mbox{m}$ dashed line) and laser
intensity ($a_0=3$ magenta pentagons, $a_0=7$ green diamonds, $a_0=10$ black squares, $a_0=14$ red circles, $a_0=20$ blue triangles).}
\label{fig: andamento}
\end{figure}

The cut-off value of the proton energy spectra (see Fig. \ref{fig: spectra evolution})
  is sharp and is related to
the maximum value of the field acting on the protons
at the beginning of the acceleration process, analogously to the ordinary TNSA case \cite{passoniNJP10}.
A summary of the maximum proton
energy obtained at time $t=166\mbox{ fs}$ in the different cases analysed by the 2D simulations
is reported in Figure  \ref{fig: andamento}-a as function of the
foam areal density (thickness in $\mum$ times $n_f/n_c$).
This comparison shows
how at a given
laser intensity, there is an ``optimal'' foam thickness for each
value of the density (for $a_0=10$, $8-12\mu\mbox{m}$ for $n_f=n_c$,
$2-4\mu\mbox{m}$ for $n_f=2n_c$ and $1-2\mu\mbox{m}$ for $n_f=4n_c$)
and, at least in the range of parameters considered, this 
corresponds to the same value of the areal density.
In Figure  \ref{fig: andamento}b) the dependence
of the maximum proton energy with respect to the foam density is reported
considering laser pulses of different powers.
 The plot shows that
at a given laser intensity, an optimal value of the foam areal density can be found 
and it is approximately proportional to the laser intensity. 
 
The increased proton energy is a direct consequence 
of the stronger electric
field arising if a foam layer is present.
The accelerating field at the
rear side of the target
is generated by the charge displacement
of the electrons which are accelerated
by the laser and propagate in the forward direction.
\begin{figure}[t]
\begin{center}
\includegraphics[width=.4\textwidth]{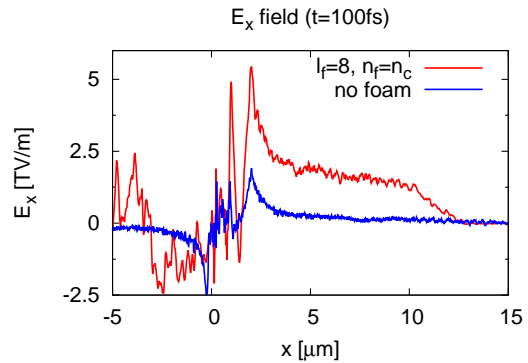}
\end{center}
\caption{Cut of the longitudinal electric field on the focal axis $z=0$
with a foam layer($n_f=1n_c$ $l_f=8\mu$m) and without,  for $a_0=10$
at $t=100$ fs  after the interaction's onset. The peak values are $E_{x,max}=5.4$TV/m
and $E_{x,max}=1.8$TV/m.}
\label{fig: Ex comparison}
\end{figure}
After the laser pulse has
been reflected by the solid layer ($t>70$fs), in presence of a foam,
the longitudinal electric field  exhibits a different shape and
a maximum value about 3 times higher if compared to the case of a bare solid foil.
Figure \ref{fig: Ex comparison} shows the line-out of the longitudinal electric field on axis for
the cases with foam ($l_f=8\mum$ $n_f=n_c$) and without for $a_0=10$.
An exponential decrease is accompanied by a nearly uniform field for
several microns from the rear surface.
Whereas the exponential decrease can be attributed to a  ``hot'' electron population
 which expands around the target
similarly to the ordinary TNSA case.
In this case, the electric field is generated due to the highly relativistic
electrons promptly escaping far away from the target, and it is strongly
different from the expression that is obtained in the assumption of a
Boltzmann equilibrium.
For foam thicknesses lower than the case of Figure \ref{fig: Ex comparison}
this ``step-like'' structure of the electric field
is still present but not as clearly distinguishable
and more similar to the case of ``ordinary'' TNSA fields.
\begin{figure}[b]
\begin{center}
\includegraphics[width=.45\textwidth]{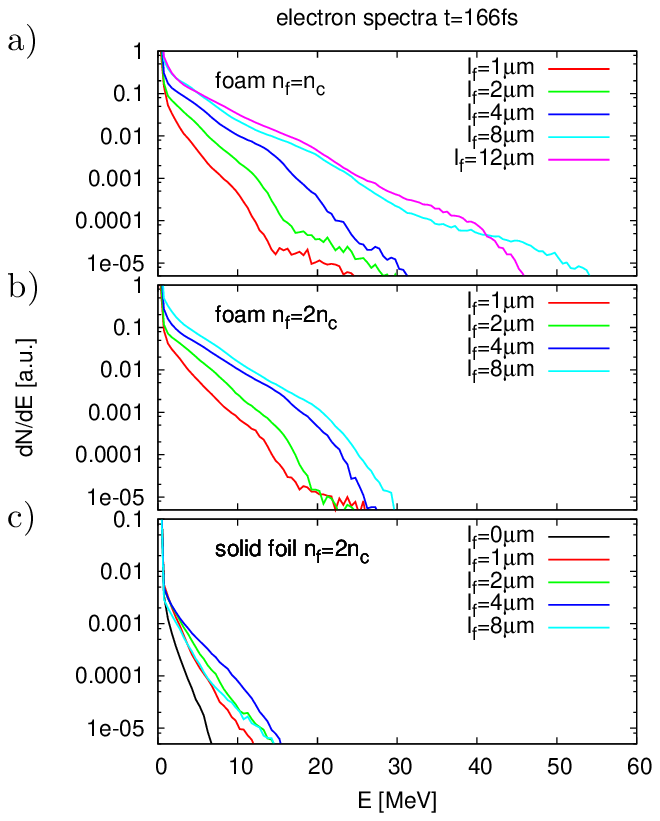}
\end{center}
\caption{Electron spectra. Electrons from the foam layer ($n_f=n_c$ a), $n_f=2n_c$ b) and electrons coming from the metal foil (c) for different simulations using a foam layer $n_f=2n_c$.}
\label{fig: e spectra}
\end{figure}

\begin{figure}[tb]
\begin{center}
\includegraphics[width=.4\textwidth]{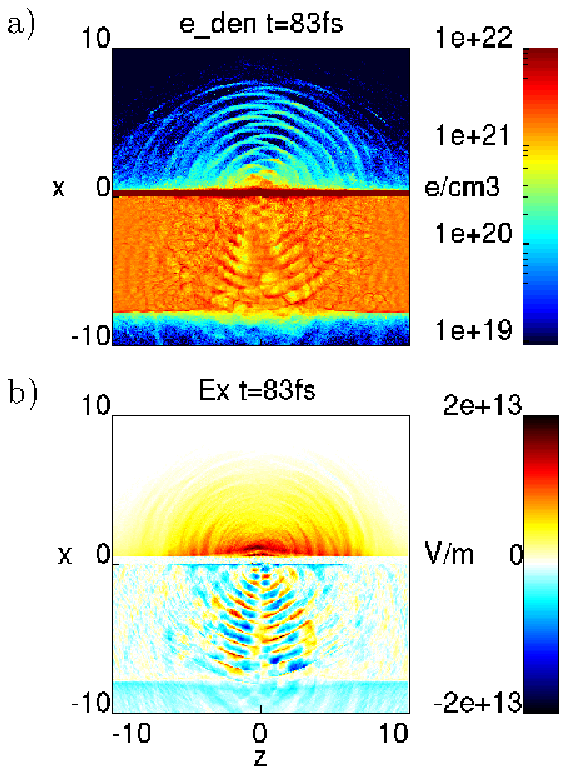}
\end{center}
\caption{$P-$polarised laser pulse ($a_0=10$) incident on a target with foam layer  $n_f=1n_c$, $l_f=8\mum$
at normal incidence: electron density (a) and longitudinal electric field (b).}
\label{fig: Ex and density}
\end{figure}
The electron spectra shown in Fig.\ref{fig: e spectra} for $a_0=10$
and foam densities $n_f=2n_c$, $n_f=n_c$ are characterized by
cut-off energies of $\simeq 30$, $\simeq 50~\mbox{MeV}$ and
rough estimates of the electrons ``temperature'' of $\simeq 9~\mbox{MeV}$
and $\simeq 10.8 \mbox{MeV}$, respectively. These values are much higher than 
the widely used ``ponderomotive'' scaling of the electron temperature
$T_p=m_ec^2(\sqrt{1+a_0^2/2}-1) \simeq 6.1m_ec^2=3.1~\mbox{MeV}$ 
for interactions with highly 
overdense plasmas, where the fast electrons have energies of the order 
of the ``quiver'' energy in vacuum, and the ``acceleration length'' is of
the order of a wavelength in such case.
The comparison shows that for the foam targets the production of fast 
electrons occurs along the volume of the foam and suggests that peculiar
mechanisms of electron acceleration are at play.
\begin{figure}[b]
\begin{center}
\includegraphics[width=.4\textwidth]{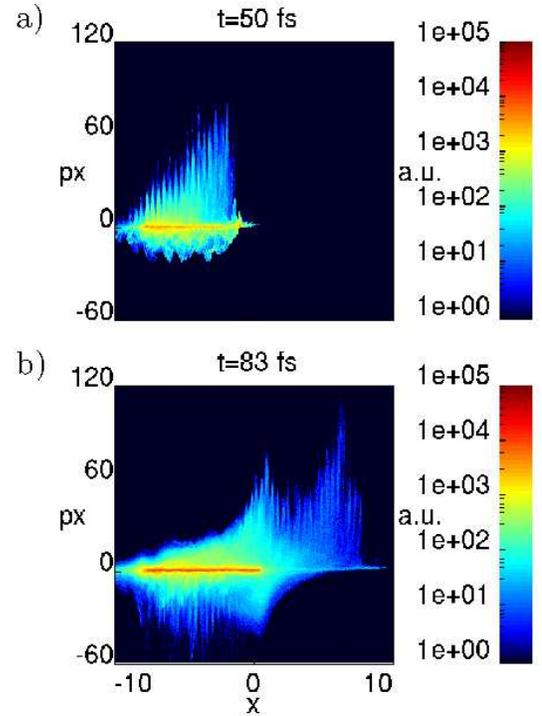}
\end{center}
\caption{Longitudinal phase space of the electrons of the foam ($n_f=1n_c$, $l_f=8\mu$m, $a_0=10$)
at two different times; $p_x$ in normalised $m_cc$ units.}
\label{fig: electron phase space}
\end{figure}
In Figures \ref{fig: Ex and density} the structures of the expanding
electrons show an antisymmetry with respect to the propagation axis.
The antisymmetry is related to the polarisation of the pulse, as it appears
for a $p$-polarisation only (i.e. for $\bfE$ in the simulation plane)
An analogous simulation has been run considering the same conditions but
with an $s-$polarised laser pulse instead. 
The ``bunching'' of the electrons is still present
but less evident and the electron density is perfectly symmetric with respect to the propagation axis.
Moreover, the electrons reach much lower energies with and energy spectrum which
extends up to about 16 MeV instead of 32 MeV with a $p-$polarised laser pulse.
As a consequence also the maximum proton energy is lower:
17 MeV instead of 30 MeV.
The comparison gives evidence of a strong contribution of the transverse 
electric field to electron acceleration.
Again, efficient penetration into the overdense plasma is necessary
in order for the coupling with the transverse field to be effective at normal pulse incidence.
In the case of a highly overdense plasma,
such that the laser penetrates only in the skin layer,
at normal incidence the coupling with the transverse electric field is inefficient and
absorption is dominated by ``$\bfJ\times\bfB$ heating'' \cite{macchiRMP11}.

The evidence of electron acceleration over the volume of the foam, bunching
of electrons and more
efficient coupling for $P$-polarization, combined with the observed 
penetration and channeling of the laser pulse inside the foam, suggests that
the dominant mechanism of fast electron generation may be similar to that 
observed in hollow microcone (funnel-like) targets, 
which allowed to obtain the highest 
proton cut-off energy experimentally observed to date \cite{gaillardPoP11}.
The mechanism, named ``direct laser-light-pressure acceleration'', relies on
the effective local grazing incidence of the laser pulse on the microcone 
walls, where the $P$-component of the electric field extracts electrons which 
are then accelerated by the combined action of ponderomotive force and 
self-generated fields, resulting in electron temperature much higher than the 
ponderomotive scaling, the increase observed in 2D simulations of 
Ref.\cite{gaillardPoP11} at $a_0=12$ from $T_p=3.9~\mbox{meV}$ to 
$9.2~\mbox{meV}$ being similar to what we observe for the foam target.
In this latter case, the penetration of the 
laser pulse inside the foam effectively yields a self-generated microcone or
funnel, providing a similar coupling at grazing incidence with the channel 
walls. The comparison with experiments and simulations in 
Ref.\cite{gaillardPoP11} may be only qualitative because of both the 
prepared microcone structure and the considerably longer and wider laser 
pulse. This implies that the total pulse energy is higher by some two orders of magnitude
than our case, which may account for the higher proton energy at similar values
of the pulse amplitude. In the present
paper we restrict to shorter pulses for reasons of numerical feasibility and 
relevance to the parameters of already proposed experiments, however it appears
of interest for further study to test the scaling of the foam-enhanced 
acceleration mechanism at higher energies of the laser pulse. In this 
perspective, from the technical point of target manufacturing the deposition of
a foam layer may be easier than engineering a hollow microcone and also 
relax alignment and pointing requirements.

\subsection{Oblique incidence}
\begin{figure}[tb]
\begin{center}
\includegraphics[width=.4\textwidth]{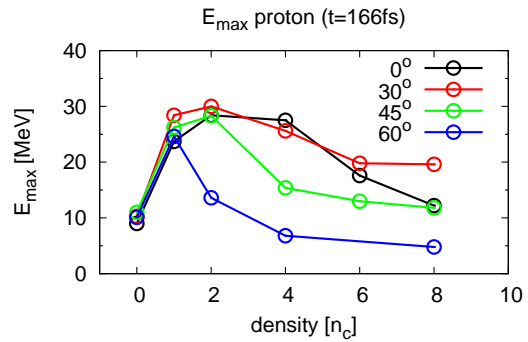}
\end{center}
\caption{Survey
of the maximum proton energy at $t=166$fs considering different target densities and angles of incidence.}
\label{fig: 45 energy}
\end{figure}

In most of the TNSA experiments, the laser pulse is focused on 
target at oblique incidence with $p-$polarisation. This increases the
 absorption and usually leads to higher proton energies.
When a foam layer is present on the irradiated side of the target, the proton energy is less
sensitive to the incidence angle.
We performed some
2D simulations considering targets with a rather thin foam layer ($l_f=2\mum$) of
different densities and a $p-$polarized laser pulse incident on target at different angles ($0\,^{\circ},\,30\,^{\circ},
\,45\,^{\circ},\,60\,^{\circ},$ $w_0=3\mum$, $a_0=10$, $\tau=25$fs).
In the case of bare target the results 
show how, despite the effective intensity ``on target'' is reduced by a factor $1/\sqrt{2}$ (for $45\,^{\circ}$),
the electron heating mechanisms at oblique incidence is
more efficient \cite{gibbon-book,brunelPRL87} leading  to higher proton energy (12 vs. 9 MeV).
In presence of a foam the proton bunch obtained  is not anymore symmetric 
with respect to the target normal and its propagation direction is slightly tilted
($\sim 2\,^{\circ}$) off axis.
One simulation has been run considering a $45\,^{\circ}$ incidence and a thick foam ($l_f=8\mu$m and $n_f=n_c$), which was
an optimal case for normal incidence, and the maximum proton energy decreased to less than 20 MeV compared to the 30 MeV of
the case at 0 degrees.
On the other hand when a thinner foam is considered
$l_f=2\mu$m and $n_f=2n_c$ (Figure \ref{fig: 45 energy})
the laser-plasma coupling is more effective
and  the resulting maximum proton energy
is comparable to the case of normal incidence.
Whereas for an angle of incidence of $30\,^{\circ}$, the results are quite similar to the case of normal incidence,
although the trend of the maximum proton energy differs for higher values of the foam density,
increasing the angle of incidence, the maximum proton energy become very sensitive to variations of the 
foam density and the optimal value is found in a narrow range of values see Figure \ref{fig: 45 energy}.

\begin{figure}[t]
\begin{center}
\includegraphics[width=.4\textwidth]{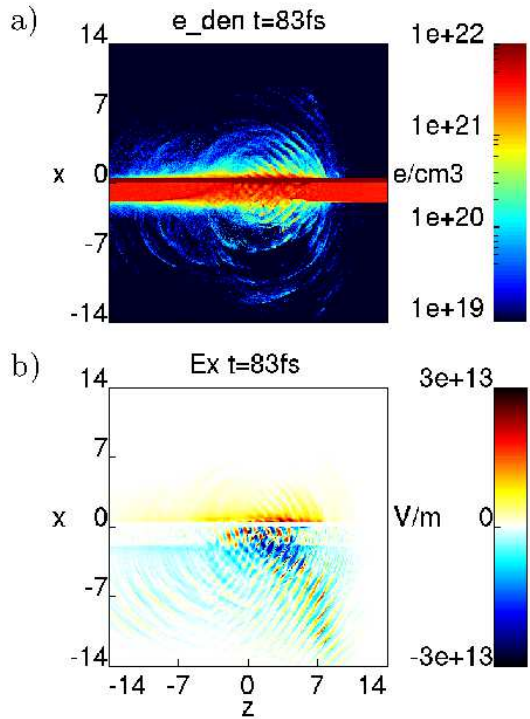}
\end{center}
\caption{$P-$polarised laser pulse ($a_0=10$) incident on a target with foam layer  $n_f=2n_c$ $l_f=2\mu$m
with a $45\,^{\circ}$ angle of incidence at time $t=83\mbox{fs}$: electron density (a) and longitudinal
electric field (b).}
\label{fig: 45 n1 l8}
\end{figure}

\section{Conclusions}
We presented an analysis of a laser driven
ion acceleration mechanism considering a
target configuration where a solid density thin plasma
is coupled with a low density layer attached on the irradiated side.
We extended the study of \textcite{nakamuraPoP10} in several directions:
performing 3D simulations, 2D simulations on a wide range of parameters
and studying the effects of oblique incidence;
moreover we discussed the mechanism of
electron acceleration and found analogies with the work of \textcite{gaillardPoP11}.
The presence of the low density layer strongly increases the energy
absorption of the target, compared to the case of a bare solid foil.
For a laser with a peak power of 32TW, the energy absorbed by the electrons goes
from about 5\% of the total laser energy for  the case without foam to over 60\%
for the cases where a foam layer with a thickness of about $4\mum$ is considered.
This translates
in a much higher number of energetic electrons with a considerably higher energy;
if a foam layer is present the electron maximum energy is more than three times
higher and the number of fast electrons is up to one order of magnitude higher
if compared to the case of a bare solid target.
These fast electrons, escaping from the rear side of the target,
build up a strong electrostatic field which easily exceeds 10TV/m and display a different.
During the interaction with the foam, the laser propagates through
the low density plasma and accelerates  to relativistic velocities
the electrons of the foam in its direction of propagation, differently than
what happens in the surface interaction with a highly overdense plasma.
 to relativistic velocities.
The comparison of 2D simulations with either $S$- and $P$-polarization
at normal incidence shows that electron acceleration is enhanced for 
$P$-polarization, which allows a coupling of the laser electric field with
the walls of the laser-drilled channel. These features and the relative 
enhancement of the electron temperature with respect to the typical 
``ponderomotive'' scaling of laser-solid interactions also suggest the 
acceleration mechanism to be similar to the one occurring in hollow microcone
targets.

The fast electrons propagate through the target and reach
the rear side creating a strong, quasi-uniform electric field.
This contribution adds to the electrostatic field arising from the
``thermal'' expansion of the hot electron population
from the solid foil, which is also irradiated by the laser
similarly to the ``pure'' TNSA regime. This configuration has been 
tested by means of PIC simulations in 3D and 2D.
The 3D simulations allowed for a more quantitative estimation of the proton acceleration and 
showed how even if a relatively thin ($l_f=2\mum$ $n_f=2n_c$) foam layer is added
on the front surface of a solid target ($l_f=0,5\mum$ $n_f=40n_c$) the energy
absorption is strongly increased and the maximum proton
energy obtained considering a 32TW laser pulse ($\tau=25$fs, $w_0=3\mum$, with a peak intensity
of $2.1\;10^{20}\wcm$), reaches 15 MeV which is more than two
times higher than the case without foam layer (6.5 MeV).
The 2D simulations, on the other hand,
 allowed for more extensive investigation of the regime
changing the target parameters
(for different thicknesses $l_f=1-12\mum$ and densities  $n_f=1-8n_c$ of the foam layer)
and the laser peak intensity over several values.
For each laser power, ranging from 3 up to 128 TW,
an optimal foam thickness can be found at a given density and the maximum
proton energy can be several times higher than case of bare solid target.
The presence of the foam proved to strongly reduce the role played by
the areal density of the solid target for the proton acceleration.
To maximize the efficiency of the process,
it is not necessary to reduce the thickness of the solid target to very low values ($\ll1\mum$)
hence avoiding the need to use targets prone to
be damaged by the pre-pulse.
If the laser irradiates the target at moderate angles ($30\,^{\circ}$) the proton energy is
not strongly affected, whereas for larger angles
the oblique incidence is considerably less efficient in producing high energy protons.
The picture arising from this investigation is richer than
the case of a pure TNSA regime and an experimental investigation
using high contrast laser might allow to study the effects of near critical
plasma with a tighter control if compared to ``pre-heated'' expanding targets.

\begin{acknowledgments}
We acknowledge the CINECA award N.HP10A25JKT-2010 (ISCRA ``TOFUSEX'' project) for the availability of
high performance computing resources and support.
We acknowledge the support of the Italian Ministry
for Education, Universities and Research specifically for the FIRB (Futuro in Ricerca)
project ``SULDIS''. This work has been performed under the auspices of the INFN ``LILIA'' project.
We also acknowledge the support of the
Ministry of Foreign Affairs (MAE), in the framework of the bilateral agreement between Italy and Japan 
for the study of the proton acceleration with interests for biomedical applications.
\end{acknowledgments}



\end{document}